\documentclass[showpacs,twocolumn,aps,prd]{revtex4-2}
\usepackage{graphicx}
\usepackage{dcolumn}
\usepackage{bm}
\usepackage{amsmath}
\usepackage{amssymb}
\usepackage{epstopdf}
\usepackage{subfigure}
\usepackage[usenames,dvipsnames]{xcolor}
\usepackage{color}
\usepackage[colorlinks,plainpages=false]{hyperref}

\begin{document}

\title{Strangelets formation in high energy heavy-ion collisions}
\author{Huai-Min~Chen,$^1$}
\email{chenhuaimin@wuyiu.edu.cn}

\author{Cheng-Jun~Xia$^{3}$}
\email{cjxia@yzu.edu.cn}
\author{Guang-Xiong~Peng$^{2,4,5}$}
\email{gxpeng@ucas.ac.cn}

\affiliation{%
    $^1$\mbox{School of Mechanical and Electrical Engineering, Wuyi University, Wuyishan 354300, China}\\
	$^2$\mbox{School of Nuclear Science and Technology, University of Chinese Academy of Sciences, Beijing 100049, China}\\
	$^3$\mbox{Center for Gravitation and Cosmology, College of Physical Science and Technology, Yangzhou University, Yangzhou 225009, China}\\
	$^4$\mbox{Theoretical Physics Center for Science Facilities, Institute of High Energy Physics, P.O. Box 918, Beijing 100049, China}\\
	$^5$\mbox{Synergetic Innovation Center for Quantum Effects and Application, Hunan Normal University, Changsha 410081, China}  }

\begin{abstract}

The properties of phase diagram of strange quark matter in equilibrium with hadronic matter at finite temperature are studied, where the quark phase and hadron phase are treated by baryon density-dependent quark mass model and hadron resonance gas model with hard core repulsion factor, respectively.
The thermodynamic conditions for the formation of metastable strange quark droplets ("strangelets") in relativistic nuclear collisions are discussed.
We obtained a rich structure of the phase diagram at finite temperature, and study the dynamical trajectories of an expanding strange fireball.
Our results indicate that the strangeness fraction $f_{s}$, perturbation parameter $C$, and confinement parameter $D$ have strong influence on the properties of phase diagram and the formation of strangelets. Consider the isentropic expansion process, we found that the initial entropy per baryon is less than or equal to 5, which gives a large probability for the formation of strangelets. Furthermore, a sufficiently large strangeness fraction $f_{\mathrm{s}}$ and one-gluon-exchange interaction and sufficiently small confinement interaction create possibilities for the formation of strangelets. On the contrary, the fireball will always complete the hadronization process when $f_\mathrm{s}=0$ or $C\geq0$ or $D^{1/2}\geq170\ \mathrm{MeV}$.

\end{abstract}

\maketitle

\section{Introduction}
\label{intro}

The phase transition between hadronic and quark matter is one of the significant and challenging fields of modern physics related to heavy-ion collisions, hybrid stars and hadronization in the early Universe.
Following Witten~\cite{Witten1984_PRD30-272} and Farhi and Jaffe~\cite{Farhi_PhysRevD.30.2379}, who proposed and studied strange quark matter, Liu and Shaw~\cite{Liu_PhysRevD.30.1137} and Greiner \emph{et al.}~\cite{Greiner_PhysRevLett.58.1825} proposed the possible creation of metastable cold strange quark matter droplets (``strangelets") in relativistic nuclear collisions.
In the early Universe, the slow expansion keeps in equilibrium the weak interactions, which make strange quarks and non-strange quarks transform into each other so that the free energy is always minimized, at the same time, the adiabatic $\beta$ equilibrium causes the "boiling off" of strange quark matter clusters created in the hadronization phase transition~\cite{Alcock_PhysRevD.39.1233}. In the case of heavy-ion collisions, strangeness must be considered as a conserved quantum number due to the suppression of weak processes by short collision time scales. Therefore, the ``$\emph{s}$-\emph{$\overline{s}$} separation" mechanism during phase transitions~\cite{LUKACS198727,Greiner_PhysRevLett.58.1825}, surface emission of hadrons~\cite{Greiner_PhysRevD.44.3517} and strangeness conservation, provide the possibility for metastable strange quark matter droplets to survive during the expansion and cooling stages of hot collision fireballs~\cite{Greiner_PhysRevLett.58.1825,Greiner_PhysRevD.44.3517}.

On the other hand, the Relativistic Heavy Ion Collider (RHIC) carried out in the early 2000s has reached the collision energy that could not be achieved in the previous heavy-ion experiments, and created matter with properties never seen at lower beam energies ~\cite{ARSENE20051,BACK200528,ADAMS2005102,ADCOX2005184}. It was supposed that the collisions could recreate the conditions of the early Universe and discover new state of matter in which quarks and gluons have been liberated from confinement~\cite{Collins_PhysRevLett.34.1353,Liu_PhysRevD.30.1137,Greiner_PhysRevLett.58.1825,Greiner_PhysRevD.38.2797,Greiner_PhysRevD.44.3517}. The state of matter is a hot Quark-Gluon Plasma (QGP). The droplets of strange quark matter (SQM), i.e., strangelets, may be formed during the cooling process of the QGP~\cite{Liu_PhysRevD.30.1137,Greiner_PhysRevLett.58.1825,Greiner_PhysRevD.38.2797,Spieles_PhysRevLett.76.1776}, which could serve as an unmistakable signature for the QGP formation in the laboratory. In reality, there are many heavy ion experiments searching for strangelets~\cite{Saito_PhysRevLett.65.2094,Shaulov_1996.4.403,Rusek_PhysRevC.54.R15,Appelquist_PhysRevLett.76.3907,Ambrosini_NPA.610.306,Sandweiss_JPG.30.S51}. However, we still know very little about the phase structure of strange matter and the thermodynamic paths followed by fireballs during expansion and cooling stages.
Therefore, it is necessary to estimate the phase transitions process in the heavy-ion collisions and provide complete and reliable thermodynamic basis and useful consistency checks for future calculations.

At present, several effective theories have been developed to study phase diagram of cold and hot strongly interacting matter, drawing important conclusions. Based on the MIT bag model, Lee and Heinz~\cite{Lee_PhysRevD.47.2068} have presented a detailed discussion of the phase structure of strange quark matter with finite strangeness and the thermodynamic conditions for the formation of strangelets in relativistic nuclear collisions, and studied the isentropic expansion process of strange quark matter systems in phase diagram. On the basis of previous study, the influence of finite volume effect on phase diagram and evolution of strangelet is further considered~\cite{He_PhysRevC.54.857}. Using a Brueckner-Hartree-Fock approach for the hadronic equation of state and a generalized MIT bag model for the quark part, Maruyama and coworkers~\cite{Maruyama_PhysRevD.76.123015} investigated the hadron-quark phase transition occurring in beta-stable matter in hyperon stars and analysed the differences between Gibbs and Maxwell phase transition constructions. Lugones and Grunfeld found that the surface tension under the MIT bag model is lower than the critical value in favor of the existence of the strangelets~\cite{GL2021prc103,GL2021prd104}. In Shao's work, they studied the influence of vector interactions on  the hadron-quark/gluon phase transition in the two-phase model, where quark matter is described by the PNJL model, and hadron matter by the nonlinear Walecka model~\cite{Shao_PhysRevD.85.114017}.

In addition, there are several other effective models describing quark matter, such as the Nambu-Jona-Lasinio (NJL) model~\cite{Mizher2010_PRD82-105016,Li2016_PRD93-054005,Maruyama2019_JCP}, quasiparticle model~\cite{Gorenstein_PhysRevD.52.5206,Peshier_PhysRevC.61.045203,Li_PhysRevD.99.043001,Zhang_PhysRevD.103.103021}, quark-cluster model~\cite{Xu_2003,Shi_2003,Xu_2010}, perturbation model~\cite{Freedman_PhysRevD.16.1130,Baluni_PhysRevD.17.2092,Fraga_PhysRevD.63.121702,Peng_EL.72.69,Xu_PhysRevD.92.025025,Xu_PhysRevD.96.063016}, and so on~\cite{Peng_PhysRevC.62.025801,Bao_EPJA.38.287,Isayev_2013,Chu_2018}. In the present paper, we apply the baryon density-dependent quark mass model considering both confinement and  first-order perturbation interactions to comprehensively study the phase diagram of quark-gluon plasma phase in equilibrium with a finite hadronic gas and analyse carefully the formation of strangelets in isentropic expansion processes. The model was proved to be thermodynamically self-consistent in the previous paper~\cite{Chen_PhysRevD.105.014011,Chen_CPC.46.055102}.

The paper is organized as follows. In Sec.~\ref{sec:Quarkphase}, we give the thermodynamic treatment and equation of state of quark phase at finite temperatures in the framework of the baryon density-dependent quark mass model. In Sec.~\ref{sec:Hadronicphase}, we consider the Hagedorn factor in hadronic phase, and give the thermodynamic treatment and equation of state. In Sec.~\ref{sec:Phase diagram and isentropic}, we present the numerical results about the properties of phase transition at finite temperature, where the phase equilibrium condition, phase diagram, and isentropic expansion process are discussed. Finally, a summary is given in Sec.~\ref{sec:sum}.

\section{The quark-gluon plasma phase}
\label{sec:Quarkphase}

The quark-gluon plasma (QGP) phase is assumed to consist of free quarks and gluons. In the framework of a baryon density-dependent quark mass model~\cite{Chen_CPC.46.055102}, the contribution of various particles to the thermodynamic potential density can be written as
\begin{equation}
\Omega_{0}=\Omega_{0}^{+}+\Omega_{0}^{-}+\Omega_{0}^{g}.
\end{equation}

The contribution of particle (+) and antiparticle ($-$) is
\begin{equation}
\Omega_{0}^{\pm}=\sum_{i}-\frac{d_{i}T}{2\pi^{2}}\int_{0}^{\infty}\ln\left[1+e^{-(\sqrt{p^{2}+m_{i}^{2}}\mp\mu_{i}^{*})/T}\right]p^{2}\mathrm{d}p.
\end{equation}

\begin{equation}
\Omega_{0}^{g}=\frac{d_{g}T}{2\pi^{2}}\int_{0}^{\infty}\ln\left[1-e^{-\sqrt{p^{2}+m_{g}^{2}}/T}\right]p^{2}\mathrm{d}p,
\end{equation}
where $i=q$ ($q=u,d,s$), $d_{q}=3(colors)\times2(spins)=6$ and $d_{g}=8(colors)\times2(spins)=16$.

In this work, we adopt a baryon density-dependent quark mass model to describe the quark mass, i.e. $m_{i}=m_{i}(n_{u},n_{d},n_{s},T)=m_{i0}+m_\mathrm{I}(n_{u},n_{d},n_{s},T)$, where $m_{i0}$, $m_\mathrm{I}$ and $n_{b}$ represent the current mass ($m_{u0}=5 \ \mathrm{MeV}, m_{d0}=5 \ \mathrm{MeV}, m_{s0}=120 \ \mathrm{MeV}$) ~\cite{PDG2022}, interaction mass and baryon density respectively. We note that the interaction mass of particles and antiparticles vary with state variables, which corresponds to strong interactions. The quark mass scaling used is
\begin{eqnarray}
m_\mathrm{i} &=&m_\mathrm{i0}+\frac{D}{n_\mathrm{b}^{1/3}}\left(1+\frac{8T}{\Lambda_{T}}e^{-{\Lambda_{T}}/{T}}\right)^{-1}\nonumber\\
             & & {} +Cn_\mathrm{b}^{1/3}\left(1+\frac{8T}{\Lambda_{T}}e^{-{\Lambda_{T}}/{T}}\right)   \label{mass}
\end{eqnarray}
with temperature scale parameter $\Lambda_{T} = 280\ \mathrm{MeV}$~\cite{Lu2016}, where $D$ corresponds to the confinement parameter and $C$ represents the strength of perturbative interactions. If $C$ takes negative values, it represents the one-gluon-exchange interaction strength~\cite{Chen2012}.

As the constant mass is replaced with an equivalent mass that varies with the environment, the relationship between the thermodynamic quantities of the free particle system is destroyed. The thermodynamic treatment is essential to ensure the thermodynamic self-consistency of the system. Starting from the free energy $\overline{F}$, the basic thermodynamic differential relationship is
\begin{equation}\label{free energy}
\mathrm{d}\overline{F}=-\overline{S}\mathrm{d}T-P\mathrm{d}V+\sum_{i}\mu_{i}\mathrm{d}{N_{i}},
\end{equation}
where $\overline{S}$, $T$, $P$, $V$, $\mu_{i}$, and $N_{i}$ correspond to the entropy, temperature, pressure, volume, chemical potential and particle number of the system, respectively. For a uniform system, the free energy density $F =\overline{F}/V $, entropy density $S =\overline{S}/V$, and particle number density $n_{i} =N_{i}/V$. The differential expression Eq.~(\ref{free energy}) becomes
\begin{eqnarray}\label{dF}
\mathrm{d}F
&=&-S\mathrm{d}T+\left(-P-F+\sum_{i}\mu_{i}n_{i}\right)\frac{\mathrm{d}V}{V} \nonumber\\
& &{}+\sum_{i}\mu_{i}\mathrm{d}n_{i}.
\end{eqnarray}

The temperature $T$, volume $V$, and quark number densities $n_{i}$ are considered as independent state variables, so the corresponding free energy density $F$ is the characteristic thermodynamic function.
However, in the thermodynamic limit, where only large volumes are considered, the thermodynamic potential is expected to scale with the volume, so that the corresponding density should be volume independent.
The free energy density takes the same form as the free particle system with the constant masses replaced by an equivalent mass, i.e.,
\begin{eqnarray}\label{F2}
F &=& F(T,\{n_{i}\},\{m_{i}\})  \nonumber\\
  &=& \Omega_{0}(T,\{\mu_{i}^{*}\},\{m_{i}\})+\sum_{i}\mu_{i}^{*}n_{i},
\end{eqnarray}
where the chemical potential $\mu_i$ is replace with the effective potential $\mu_i^*$. The corresponding differential relation is
\begin{eqnarray}\label{ddF}
\mathrm{d}F &=& \frac{\partial\Omega_{0}}{\partial T}\mathrm{d}T+\sum_{i}\left(\frac{\partial\Omega_{0}}{\partial \mu_{i}^{*}}\mathrm{d}\mu_{i}^{*}+\mu_{i}^{*}\mathrm{d}n_{i}+n_{i}\mathrm{d}\mu_{i}^{*}\right)\nonumber\\
          & &{} +\sum_{i}\frac{\partial\Omega_{0}}{\partial m_{i}}\left(\sum_{j}\frac{\partial m_{i}}{\partial n_{j}}\mathrm{d}n_{j}+\frac{\partial m_{i}}{\partial T}\mathrm{d}T\right) \nonumber\\
          &=& \left[
      \frac{\partial \Omega_0}{\partial T}
      + \sum_i \frac{\partial \Omega_0}{\partial m_i} \frac{\partial m_i}{\partial T}
\right]\mbox{d}T
\nonumber \\
&&
+
\sum_i \left[ \mu_i^* +  \sum_j\frac{\partial \Omega_0}{\partial m_j}\frac{\partial m_j}{\partial n_i}
\right] \mbox{d} n_i.
\end{eqnarray}

By comparing Eq.~(\ref{dF}) with Eq.~(\ref{ddF}), we have
\begin{eqnarray}
S^{Q} &=& -\frac{\partial\Omega_{0}}{\partial T}-\sum_{i}\frac{\partial m_{i}}{\partial T}\frac{\partial\Omega_{0}}{\partial m_{i}}, \label{7}
\end{eqnarray}
\begin{eqnarray}
P^{Q} &=&  -F+\sum_i \mu_{i} n_i^{Q} , \label{8}
\end{eqnarray}
\begin{eqnarray}
\mu_{i} &=& \mu_i^* +   \sum_j\frac{\partial \Omega_0}{\partial m_j}\frac{\partial m_j}{\partial n_i}, \label{9}
\end{eqnarray}
where the up index Q is used to label the quark phase. Combining Eq.~(\ref{8}) and $E=F-TS$ shows that Euler equation $E = TS - P + \sum_{i} \mu_{i} n_{i}$ is verified.

The energy density is given by
\begin{eqnarray}
E = F+TS=\Omega_{0}+\sum_{i}\mu_{i}^{*}n_{i}^{Q}+TS. \label{11}
\end{eqnarray}

By $\partial F/\partial \mu_{i}^{*}=\partial \Omega_{0}/\partial \mu_{i}^{*} + n_{i}=0$, the particle number densities $n_{i}^{Q}=n_{i}^{+}-n_{i}^{-}$ are obtained by
\begin{eqnarray}
n_{i}^{\pm} = -\frac{\partial\Omega_{0}^{\pm}}{\partial \mu_{i}^{*}}. \label{10}
\end{eqnarray}

Considering the contribution of gluons to the system, we need to know the effective mass of gluons. Recently, Bors\'{a}nyi $\emph{et al.}$ gave 48 pressure values from the lattice simulation~\cite{Borsanyi.Endrodi.ea1-22JHEP}. Based on pressure in lattice data, we could describe the gluon mass according to the fast convergence expression of QCD coupling. By the corresponding 48 pressure values, we use the least square method to obtain the most effective fitting results. Here, we define the scaled temperature as $x= T/T_{c}$, where $T_{c}$ is the critical temperature. At $T<T_{c}$, the expression of gluon's equivalent mass is
\begin{equation}\label{32}
  \frac{m_{g}}{T}=\sum_{i}a_{i}x^{i}=a_{0}+a_{1}x+a_{2}x^{2}+a_{3}x^{3},
\end{equation}
where expansion coefficients $a_{0}=67.018$, $a_{1}=-189.089$, $a_{2}=212.666$, $a_{3}=-83.605$.
At $T> T_{c}$, the expression of gluon's equivalent mass is
\begin{equation}\label{33}
  \frac{m_{g}}{T}=\sum_{i}b_{i}\alpha^{i}=b_{0}+b_{1}\alpha+b_{2}\alpha^{2}+b_{3}\alpha^{3},
\end{equation}
where $\alpha$ is the strong coupling constant and expansion coefficients $b_{0}=0.218$, $b_{1}=3.734$, $b_{2}=-1.160$, $b_{3}=0.274$. As is well know, the QCD coupling $\alpha$ is running and depends on the solution of the renormalization-group equation for the coupling. Recently, we have solved the renormalization group equations for the QCD coupling by a mathematically strict way and obtained a fast convergence expression of $\alpha$~\cite{Chen2022IJMPE}. Here, we only take the leading order term, i.e
\begin{eqnarray}\label{34}
\alpha=\frac{\beta_0}{\beta_0^2\ln(u/\Lambda)+\beta_1\ln\ln(u/\Lambda)},
\end{eqnarray}
where the beta coefficients $\beta_0={11}/{2}-{N_f}/{3}$, $\beta_1={51}/{4}-{19N_f}/{12}$, and number of flavors $N_{f}=3$. The renormalization scale varies linearly with temperature as ${u}/{\Lambda}=c_0+c_1x$, $c_{0}=1.054$, $c_{1}=0.479$.

\begin{figure}[htbp]
\centering
\includegraphics[width=8.5cm]{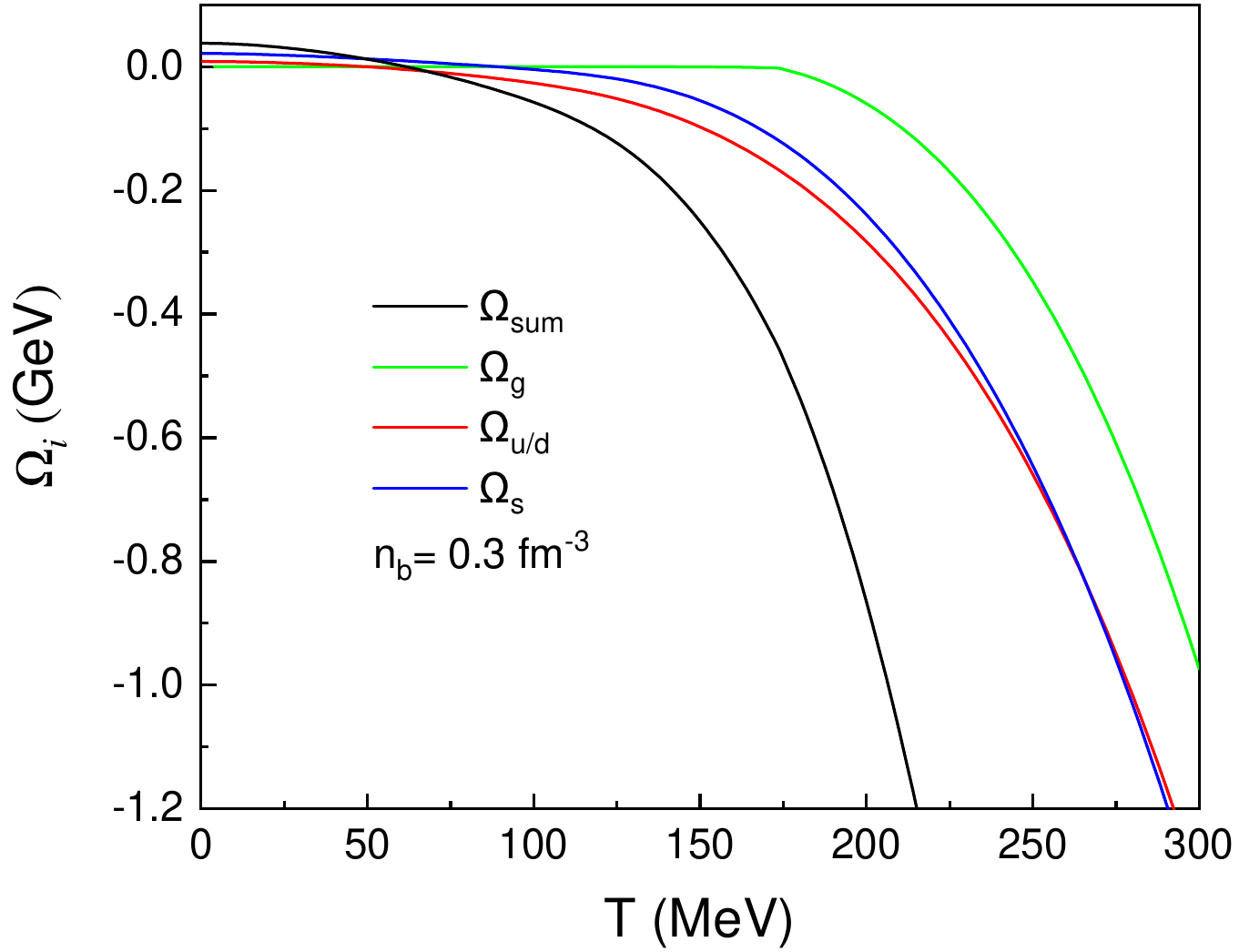}
\caption{Temperature dependence of contribution of particle species to the thermodynamic potential of light quarks (u/d), massive strange quark (s), and gluon (g) at fixed $n_b$. The ``sum" stands for the total contributions.}
\label{Fig1}
\end{figure}

To get an impression of the gluon introduced here, we have plotted the temperature dependence of contribution of particle species to the thermodynamic potential in Fig.~\ref{Fig1}. It can be seen that they are all decreasing functions of the temperature. The gluon contribution of the thermodynamic potential vanishes at zero temperature, while for quarks there are finite contribution even at zero temperature. We see that below the critical temperature, the contribution of the thermodynamic potential of gluons is very small, while above the critical temperature, the thermodynamic potential of gluons can rapidly increase. However, the contribution of gluons to the total thermodynamic potential is relatively small, while the contribution of quarks is relatively large.

\section{Hadronic phase }
\label{sec:Hadronicphase}

We consider hadronic phase as a weakly interacting mixed gas of strange hadrons $K^+, K^0, \Lambda, \Sigma, \Xi, \Omega$, non-strange hadrons $\pi, \eta, N, \Delta$(1232), and their anti-particles. According to the quark component of various hadrons, the chemical potential of hadrons is composed of quark chemical potential as follows
\begin{eqnarray}
{\mu}_i=\sum_q(n^i_q-n^i_{\bar{q}})
\mu_q,
\end{eqnarray}
where $i$ and $q$ represent the hadronic species and quark flavor respectively. $(n_{q}^{i}-n_{\bar{q}}^{i})$ is the net number of the quark q for $i$-th baryon.

Based on Bose$-$Einstein and Fermi$-$Dirac statistics, the expressions of the thermodynamic quantities for \textit{i}-th noninteracting hadrons are
\begin{equation}
\varepsilon_i^\mathrm{pt}=\frac{d_i}{2 \pi^2}\int_0^\infty \frac{p^2 {\varepsilon}_i}
 {e^{({\varepsilon}_i-{\mu}_i)/T}\pm1}+\frac{p^2 {\varepsilon}_i}
 {e^{({\varepsilon}_i+{\mu}_i)/T}\pm1}d p,
\end{equation}
\begin{equation}
P^\mathrm{pt}_i = \frac{d_i}{6\pi^2}\int_0^\infty\frac{p^4}{{\varepsilon}_i(e^{({\varepsilon}_i-{\mu}_i)/T}\pm1)}
+\frac{p^4}{{\varepsilon}_i(e^{({\varepsilon}_i+{\mu}_i)/T}\pm1)}d p,
\end{equation}
\begin{equation}
n^\mathrm{pt}_i=\frac{d_i}{2 \pi^2}\int_0^\infty \frac{p^2}
{e^{({\varepsilon}_i-{\mu}_i)/T}\pm1}+\frac{p^2}
{e^{({\varepsilon}_i+{\mu}_i)/T}\pm1} d p,
\end{equation}
\begin{eqnarray}
S^\mathrm{pt}_i&=&\pm\frac{ d_i}{2\pi^2} \int_0^{\infty}
  \left\{
 \ln[1\pm e^{-({\epsilon}_{i}-{\mu}_i)/T}]\pm\frac{({\epsilon}_{i}-{\mu}_i)/T}{e^{({\epsilon}_{i}-{\mu}_i)/T}\pm
1}
  \right.\nonumber\\
&& \left.
+\ln[1\pm e^{-({\epsilon}_{i}+{\mu}_i)/T}]\pm\frac{({\epsilon}_{i}+{\mu}_i)/T}{e^{({\epsilon}_{i}+{\mu}_i)/T}\pm
1}
  \right\}
  p^2\mathrm{d}p,
\end{eqnarray}
where ${\epsilon}_i=\sqrt{p^2+{m}_i^2}$, and the upper and the lower operation symbol  denotes the fermions and bosons respectively. The parameter $d_{i}$ represent degeneracy factors $d_{i}=$  spin $\times$ isospin. Naturally, the total energy density, pressure, and baryon number density for the hadronic phase are
\begin{eqnarray}
\varepsilon^\mathrm{pt}&=&\sum_i \varepsilon_i^\mathrm{pt},\\
P^\mathrm{pt}&=&\sum_iP^\mathrm{pt}_i,\\
 n_{\mathrm{b}}^\mathrm{pt}&=&\sum_ib_{i}n^\mathrm{pt}_i.
\end{eqnarray}

Here, a proper volume correction of point-like hadron is used to consider the hard core repulsion, which is known as the Hagedorn correction factor~\cite{Lee_PhysRevD.47.2068,Hagedorn1980PLB,Wen2007CITP}. Then, the energy density, pressure, baryon number density, and entropy density are modified , i.e.,
\begin{eqnarray}
E^\mathrm{H}&=&\frac{1}{1+\varepsilon^\mathrm{pt}/{4 B}}\sum_i
\varepsilon_i^\mathrm{pt},\\
P^\mathrm{H}&=&\frac{1}{1+\varepsilon^\mathrm{pt}/{4 B}}\sum_i
P^\mathrm{pt}_i,\\
 n_{\mathrm{b}}^\mathrm{H}&=&\frac{1}{1+\varepsilon^\mathrm{pt}/{4 B}}\sum_ib_i
n^\mathrm{pt}_i,\\
S^\mathrm{H}&=&\frac{1}{1+\varepsilon^\mathrm{pt}/{4
B}}\sum_iS^\mathrm{pt}_i,
\end{eqnarray}
where $b_i$ corresponds to the baryon number of \textit{i}-th hadron. The factor $(1+\varepsilon^\mathrm{pt}/{4 B})^{-1}$ is the proper volume correction and limits the energy density to $4B$, where the bag constant $B^{1/4}=180 \ \mathrm{MeV}$~\cite{PDG2022}.

The number density of strange quarks is
\begin{eqnarray}
n_\mathrm{s}^\mathrm{H}=\frac{1}{1+\varepsilon^\mathrm{pt}/{4
B}}\sum_i s_i n^\mathrm{pt}_i,
\end{eqnarray}
where $s_i$ is the strange valence quark number of \textit{i}-th hadron.

Note that in an ideal HRG mode~\cite{Bhattacharyya2016EPL115}, the hadrons are assumed to be point-like particles with no interaction between them. However, this assumption is very simplistic and fails to describe the lattice QCD (LQCD) data at temperatures above $T\sim150\ \mathrm{MeV}$ The validity region of the hadron resonance gas (HRG) model could be extended to higher temperature by introducing the interaction effect. The excluded volume (EV) effect is the simplest way to implement the interacting HRG, the short-range repulsive interaction was modeled via a hard-core correction following the thermodynamically consistent way as developed in Ref.~\cite{Hagedorn1980PLB}.

Further improvement was made considering Lorentz contracted hard core potentials in hadron gas models~\cite{Bugaev2000PLB485}. It was found that at higher temperatures, the Lorentz contraction effects are stronger for light particles such as pions and make their effective excluded volume smaller than that of heavy ones. At smaller temperatures ($T\sim100\ \mathrm{MeV}$), the pion density is small and excluded volume corrections are unimportant~\cite{Pal _NPA.A1010.122177}. One of the most successful improvement to the model which explains the LQCD results is the van der Waals hadron resonance gas (VDWHRG) model. The VDWHRG model effectively explains the LQCD data up to $T\sim180\ \mathrm{MeV}$~\cite{Bhagyarathi _PhysRevD.108.074028}. van der Waals interaction does play a crucial role in the hadronic systems at high temperatures~\cite{Vovchenko _PhysRevLett.118.182301}.

Our article mainly studies the structure of phase diagrams and the evolution of isentropic expansion processes. Since the temperature during the isentropic expansion process is generally below $100\ \mathrm{MeV}$, we expect the volume correction of meson is not significant. Therefore, the current article simply considers the EV effect. In the future work, we plan to consider more realistic models.

\section{Phase diagram and isentropic expansion process at finite temperature}
\label{sec:Phase diagram and isentropic}

\begin{figure*}[htbp]
\centering
\includegraphics[width=15cm]{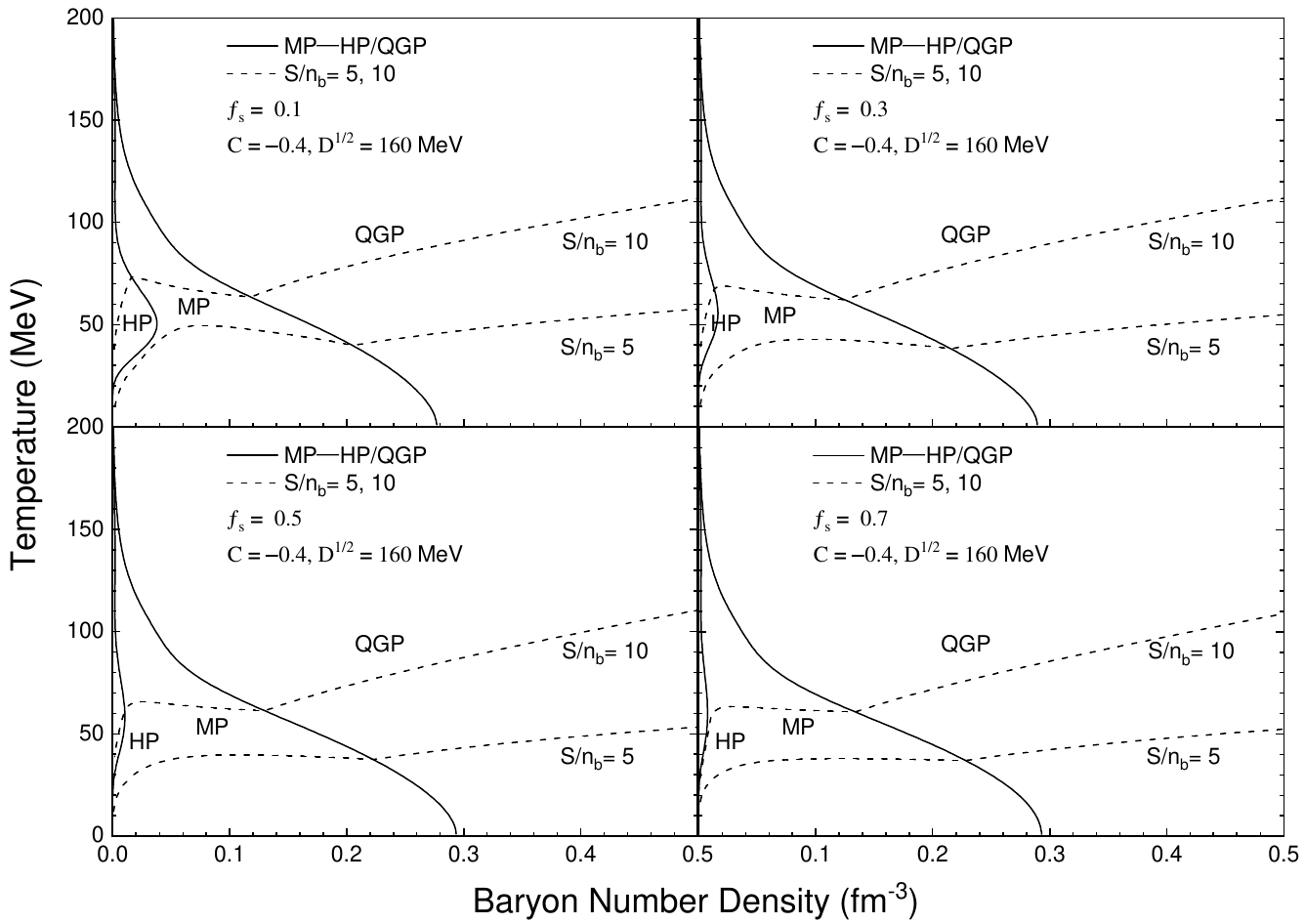}
\caption{The phase diagram and isentropic expansion process have been shown at different fixed strangeness fraction.}
\label{Fig2}
\end{figure*}

We consider the isolated system with finite strangeness that undergoes a first order phase transition from the hadronic phase to the QGP phase. In the equilibrium phase diagram, eigen quantities satisfy Gibbs equilibrium conditions, i.e., chemical equilibrium $\mu_i^\mathrm{Q}=\mu_i^\mathrm{H}$,  mechanic equilibrium $P^\mathrm{Q}=P^\mathrm{H}$, and thermodynamic equilibrium $T^\mathrm{Q}=T^\mathrm{H}$. In the isolated system, the total net baryon number in the system is kept a constant. In addition, the total net strangeness in the compact system is conserved since the collision in heavy-ion collisions is too short to establish flavor equilibrium~\cite{He_PhysRevC.54.857}. The strangeness fraction is defined as
\begin{eqnarray}
f_\mathrm{s}=n_\mathrm{s}^\mathrm{tot}/n_\mathrm{b}^\mathrm{tot}.
\end{eqnarray}
We consider that the strangeness fraction is in the range $0\leq f_\mathrm{s}<3$. The system maintains a fixed strangeness fraction, resulting in a smooth variation of the chemical potentials  during the conversion from hadronic matter to QGP.
Here, we will not conduct a complete dynamic study on the expansion of fireballs, including the effects of surface evaporation which could change the strangeness and entropy density of the system [2, 5, 6] and the dynamics of the freeze-out process. On the contrary, we will attempt to obtain some qualitative insights by assuming a smooth hydrodynamic dynamics expansion at strangeness fraction $f_\mathrm{s}$ and constant entropy $\overline{S}/A$.

The phase transitions occur through a mixed phase. For the quark phase, it is difficult for the system to achieve mechanical equilibrium since the quark mass will become infinite when $n_\mathrm{b}^\mathrm{Q}\rightarrow0$. Referring to the method used by He et al.~\cite{He_PhysRevC.54.857}, we define a ratio of the hadronic phase volume to the total volume as $\alpha = V^\mathrm{H}/V^\mathrm{tot}$. $\alpha=0$ and $\alpha=1$ correspond to the beginning and end of hadronization respectively, and then we obtain the boundary between mixed phase and hadronic or quark phase. Similarly, we define $n_\mathrm{b}^\mathrm{Q} = N_\mathrm{b}^\mathrm{Q}/V^\mathrm{Q}$, $n_\mathrm{b}^\mathrm{H} = N_\mathrm{b}^\mathrm{H}/V^\mathrm{H}$ to represent the baryon number density in the quark and hadronic phases. Generally, common light quarks chemical potentials $\mu_{u}=\mu_{d}$ are assumed by isospin symmetry~\cite{He_PhysRevC.54.857,Wen2007CITP}. According to the Gibbs conditions and baryon/strangeness density conservation condition, the equilibrium phase diagram satisfies
\begin{eqnarray}
& P^\mathrm{Q}(T,\mu_q,\mu_\mathrm{s},n_\mathrm{b}^\mathrm{Q})
 =P^\mathrm{H}(T,\mu_q,\mu_\mathrm{s})\label{QMDTDeq1},  & \\
&
n_\mathrm{b}^\mathrm{tot}
 =n_\mathrm{b}^\mathrm{Q}
   (T,\mu_q,\mu_\mathrm{s},n_\mathrm{b}^\mathrm{Q})(1-\alpha)
  +n_\mathrm{b}^\mathrm{H}(T,\mu_q,\mu_\mathrm{s})\alpha, & \\
&
n_\mathrm{s}^\mathrm{tot}
 =n_\mathrm{s}^\mathrm{Q}(T,\mu_\mathrm{s},n_\mathrm{b}^\mathrm{Q})(1-\alpha)
  +n_\mathrm{s}^\mathrm{H}(T,\mu_q,\mu_\mathrm{s})\alpha\label{QMDTDeq3},  & \\
&
S^\mathrm{tot}
=S^\mathrm{Q}(T,\mu_q,\mu_\mathrm{s},n_\mathrm{b}^\mathrm{Q})(1-\alpha)
 +S^\mathrm{H}(T,\mu_q,\mu_\mathrm{s})\alpha\label{entroQMDTD}.   &
\label{QMDTDeq2}
\end{eqnarray}

In the framework of the thermodynamic treatment method of strange quark matter and hadronic matter given in Sec.~\ref{sec:Quarkphase} and ~\ref{sec:Hadronicphase}, we could obtain the phase structure by solving Eqs.~(\ref{QMDTDeq1})-(\ref{QMDTDeq3}). Considering Eq.~(\ref{entroQMDTD}), we will get the isentropic expansion process with baryon density-dependent quark mass model.
\begin{figure*}[htbp]
\centering
\includegraphics[width=15cm]{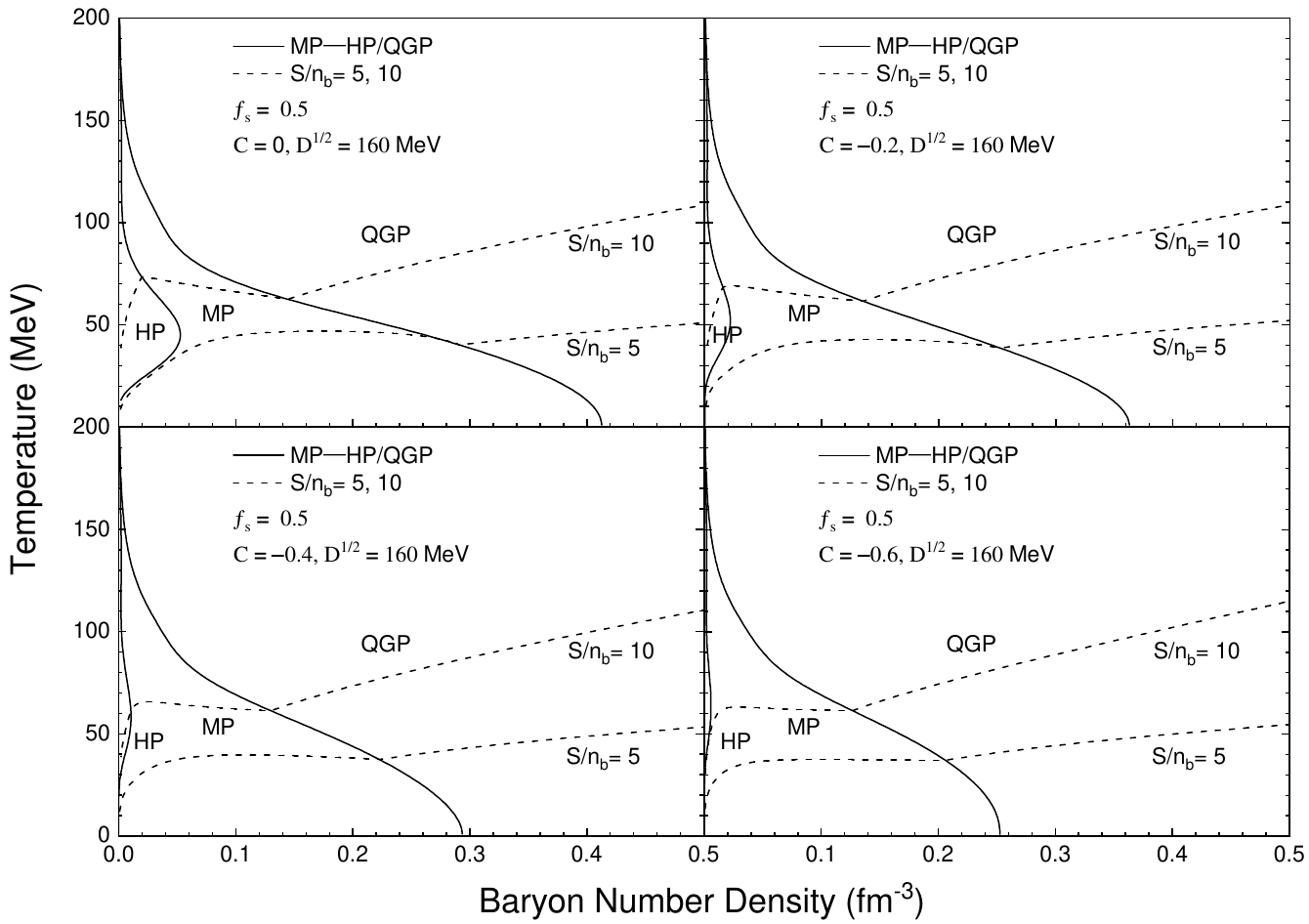}
\caption{The phase diagram and isentropic expansion process have been shown under different perturbation parameters $C$.}
\label{Fig3}
\end{figure*}

At fixed strangeness fraction $f_{\mathrm{s}}$, we get the phase diagram in Fig.~\ref{Fig2} by solving Eqs.~(\ref{QMDTDeq1})-(\ref{QMDTDeq3}). The dashed curves obtained at $\alpha$=0 represents the boundary between the quark-gluon phase and the mixed phase. When the value of $\alpha$ approaches 1, we obtain the boundary between the hadron phase and the mixed phase, which is indicated by solid curves. From the phase diagram, we can see that the quark phase is on the right side of each diagram, the hadron phase is on the left side, and the mixed phase is in the middle. The strangeness fraction has a significant impact on the boundary between the hadron phase and the mixed phase. As the strangeness fraction increases, the area of the hadron phase dramatically decrease, and the boundary curve between the hadron phase and the mixed phase will approach the temperature axis. The boundary between the quark phase and the mixed phase does not vary significantly with the strangeness fraction, but only slightly expands to the right. Therefore, the area of the mixed phase is constantly expanding. Furthermore, we found a narrow high temperature and low density mixed phase region in the phase diagram, which implies the possibility of forming strangelets at HICs. In addition, our research shows that a large strangeness fraction is beneficial for the formation of strangelets during the process of quark-hadron phase transition. This is consistent with the conclusion of previous model studies such as QMDTD model~\cite{Wen2007CITP}.

Next, we discuss the isentropic expansion process with the initial entropies per baryon under fixed strangeness fraction, in which the system is adiabatic and total entropy is conserved. Based on Eqs.~(\ref{QMDTDeq1})-(\ref{QMDTDeq2}), We obtained the expansion trace under entropy conservation.
The isentropic expansion process under different strangeness fraction $f_\mathrm{s}$ are also shown in Fig.~\ref{Fig2}. The two dashed curves represent the isentropic expansion trace, and the initial entropy per baryon is 5 and 10, respectively. We find that a high initial entropies per baryon will prevent the occurrence of strangelets at the final stage of evolution, and the fireball will always complete the hadronization process. At the strangeness fraction $f_\mathrm{s}= 0.1, 0.3, 0.5$, the initial entropy per baryon is about 5, which is beneficial for the formation of strangelets. Compared with the MIT bag model ~\cite{Wen2007CITP,He_PhysRevC.54.857,Lee_PhysRevD.47.2068}, the baryon density-dependent quark mass model predicts a similar isentropic expansion trajectory for the formation of strangelets.
However, the difference is that the mixed phase in the baryon density-dependent quark mass model has a narrow region of mixed phase at high temperature and low density.
In case of high entropy, the isentropic curve for the mixed phase is shorter, and the reheating effect of baryon density-dependent quark mass model is more significant than that of the bag model.
Moreover, as the strangeness fraction $f_{\mathrm{s}}$ decreases, the area of hadron phase will expand and the area of mixed phase will shrink in the phase diagram, the reheating effect is more significant, reducing the chances for the formation of strangelets so that fireball will make it easier to complete the hadronic process.
\begin{figure*}[htbp]
\centering
\includegraphics[width=15cm]{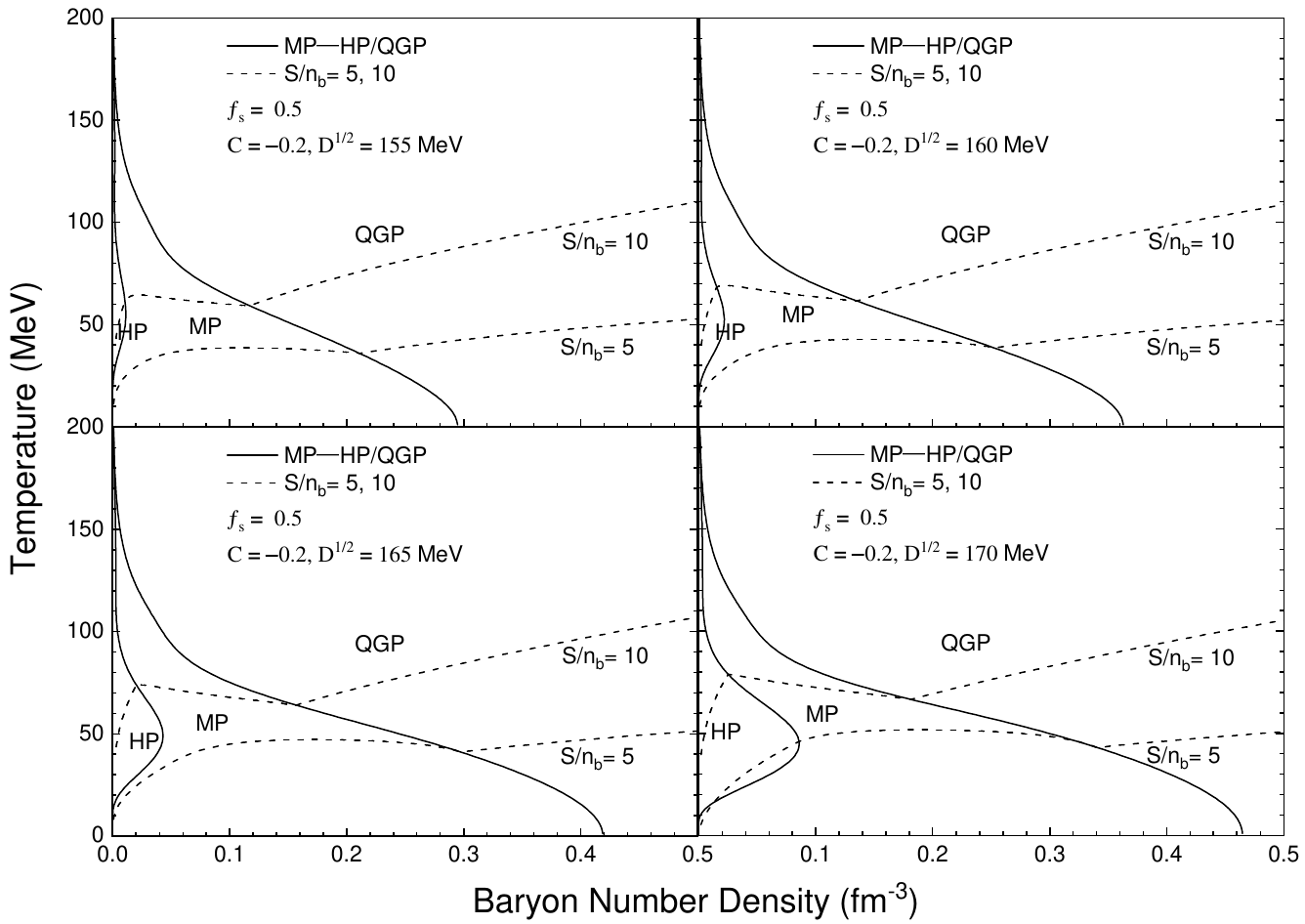}
\caption{The phase diagram and isentropic expansion process have been shown under different confinement parameter $D$.}
\label{Fig4}
\end{figure*}

In Fig.~\ref{Fig3}, we present phase diagrams adopting different strengths for one-gluon-exchange interactions. We can see that the boundary curve between the hadron phase and the mixed phase will expand to the right, the boundary between the quark phase and the mixed phase also expands to the right and has a more significant impact as the one gluon-exchange interaction strength $C$ decreases. In other words, the area of the mixed phase and hadron phase is continuously decreasing with the strength of one-gluon-exchange interactions. Therefore, a larger one gluon-exchange interaction strength $C$ is conducive to the formation of strangelets.

At the same time, we consider the isentropic expansion process under different one-gluon-exchange interaction strength in Fig.~\ref{Fig3}. Based on the analysis of phase diagrams with different strangeness fraction, we take $f_\mathrm{s}$=0.5.
At $f_\mathrm{s}=0.5$, all systems with $\mathrm{S}/\mathrm{n}_{\mathrm{b}}\leq5$ follow expansion trajectories and are trapped in the mixed phase.
As the one-gluon-exchange interaction strength increases, the area of hadronic phase and mixed phase in the phase diagram will be reduced, and the reheating effect will not change significantly. The fireball will not complete the hadronization process and remains within the mixed phase, creating the possibility for the formation of strangelets. When $C$ takes a positive value, it represents perturbation interaction, and the fireball will always complete hadronization process.

The phase diagrams under different confinement parameters is indicated in Fig.~\ref{Fig4}. As the confinement parameter $D$ increases, the boundary curve between the hadron phase and the mixed phase shifts to the right, and the boundary between the quark phase and the mixed phase also shifts to the right, where the mixed phase covers a larger density range. Therefore, a smaller confinement parameter $D$ is beneficial for the formation of strangelets.

In addition, the phase diagram and isentropic expansion process under different confinement parameter $D$ are shown in Fig.~\ref{Fig4}. At $f_\mathrm{s}=0.5$, all systems with $\mathrm{S}/\mathrm{n}_{\mathrm{b}}\leq5$ follow expansion trajectories and are trapped in the mixed phase.
As the confinement parameter $D$ increases, the area of hadron phase and mixed phase will expand in the phase diagram, the reheating effect is more significant, decreasing the possibility for the formation of strangelets. Until $D^{1/2}=170\ \mathrm{MeV}$, the isentropic expanding fireball will complete hadronization process.

As shown in Figs.~\ref{Fig2}-\ref{Fig4}, even if the baryon density is very small, some isentropic trajectories of $\mathrm{S}/\mathrm{n}_{\mathrm{b}}$ do not enter the hadronic phase, i.e., the system will not be completely hadronized. This means that the strange quark matter can actually survive the hadronization and cooling process. As the temperature decreases, QGP cools down, giving rise to surviving (meta)stable cold strangelets. This is consistent with the conclusion of reference~\cite{Lee_PhysRevD.47.2068}.

\section{SUMMARY}
\label{sec:sum}

We have considered the phase structure of strange matter, revealing the rich structure of the quark-hadron phase diagram, providing a comprehensive and reliable thermodynamic basis for the dynamic study of the creation of strange quark matter in relativistic nuclear collisions.

The phase diagram of strange quark matter in equilibrium with hadronic matter is systematically studied at finite temperature within baryon density-dependent quark mass model and hard core repulsion factor. Based on the Gibbs equilibrium conditions, we studied the effects of the strangeness fraction $f_{\mathrm{s}}$, quark confinement and first-order perturtabative interactions on the phase diagram, the isentropic expansion process and the formation of strangelets.
In the context of isentropic and hydrodynamic expansion, the formation of cold strangelets requires a sufficiently small confinement interaction $D$ and large one-gluon-exchange interaction $C$, and an expanding fireball with either a low specific entropy $\mathrm{S}/\mathrm{n}_{\mathrm{b}}$ or a large strangeness fraction $f_{\mathrm{s}}$. On the contrary, the fireball will always complete the hadronization process when $f_\mathrm{s}=0$ or $C\geq0$ or $D^{1/2}\geq170\ \mathrm{MeV}$.
Therefore, a sufficiently large strangeness fraction $f_{\mathrm{s}}$ and one-gluon-exchange interaction $C$ and sufficiently small confinement interaction $D$ create possibilities for the formation of strangelets. Futhermore, we found that the initial entropy per baryon is less than or equal to 5, the fireball will not complete the hadronization process and remains within the mixed phase, creating the possibility for the formation of strangelets.

\appendix

\section*{ACKNOWLEDGMENTS}

The authors would like to thank support from NSFC (Nos. 11135011, 12275234, 12375127) and the national SKA programe (No. 2020SKA0120300).


\begin{thebibliography}{99}

\bibitem{Witten1984_PRD30-272}
E. Witten,
Phys. Rev. D 30, 272 (1984).

\bibitem{Farhi_PhysRevD.30.2379}
E. Farhi and R. L. Jaffe,
Phys. Rev. D 30, 2379 (1984).

\bibitem{Liu_PhysRevD.30.1137}
H. C. Liu and G. L. Shaw,
Phys. Rev. D 30, 1137 (1984).

\bibitem{Greiner_PhysRevLett.58.1825}
 C. Greiner and  P. Koch and H. St\"ocker,
Phys. Rev. Lett. 58, 1825 (1987).

\bibitem{Alcock_PhysRevD.39.1233}
 C. Alcock and  A. Olinto,
Phys. Rev. D 39, 1233 (1989).

\bibitem{LUKACS198727}
 B. Luk\'{a}cs and J. Zim\'{a}nyi and N. L. Balazs,
Phys. Lett. B 183, 27 (1987).

\bibitem{Greiner_PhysRevD.44.3517}
C. Greiner and H. St\"ocker,
Phys. Rev. D 44, 3517 (1991).

\bibitem{ARSENE20051}
I. Arsene {\it et al.} [BRAHMS Collaboration],
Nucl. Phys. A 757, 1 (2005).

\bibitem{BACK200528}
B. B. Back {\it et al.} [PHOBOS Collaboration],
Nucl. Phys. A 757, 28 (2005).

\bibitem{ADAMS2005102}
J. Adams {\it et al.} [STAR Collaboration],
Nucl. Phys. A 757, 102 (2005).

\bibitem{ADCOX2005184}
K. Adcox {\it et al.} [PHENIX Collaboration],
Nucl. Phys. A 757, 184 (2005).

\bibitem{Collins_PhysRevLett.34.1353}
J. C. Collins and M. J. Perry,
Phys. Rev. Lett. 34, 1353 (1975).

\bibitem{Greiner_PhysRevD.38.2797}
C. Greiner and D. H. Rischke and  H. St\"ocker and P. Koch,
Phys. Rev. D 38, 2797 (1988).

\bibitem{Spieles_PhysRevLett.76.1776}
 C. Spieles and L. Gerland and  H. St\"ocker and C. Greiner and C. Kuhn and J. P. Coffin,
Phys. Rev. Lett. 76, 1776 (1996).

\bibitem{Saito_PhysRevLett.65.2094}
T. Saito and Y. Hatano and Y. Fukada and  H. Oda,
Phys. Rev. Lett. 65, 2094 (1990).

\bibitem{Shaulov_1996.4.403}
S. B. Shaulov,
Acta Phys. Hung. A 4, 403 (1996).

\bibitem{Rusek_PhysRevC.54.R15}
A. Rusek and B. Bassalleck and A. Berdoz and T. B\"urger and M. Burger and R. E. Chrien {\it et al.},
Phys. Rev. C 54, R15 (1996).

\bibitem{Appelquist_PhysRevLett.76.3907}
G. Appelquist and C. Baglin and J. Beringer and C. Bohm and K. Borer and A. Bussi\`ere  {\it et al.},
Phys. Rev. Lett. 76, 3907 (1996).

\bibitem{Ambrosini_NPA.610.306}
G. Ambrosini  {\it et al.},
Nucl. Phys. A 610, 306 (1996).

\bibitem{Sandweiss_JPG.30.S51}
J. Sandweiss,
J. Phys. G: Nucl. Part. Phys. 30, S51 (2004).

\bibitem{Lee_PhysRevD.47.2068}
K. S. Lee and U. Heinz,
Phys. Rev. D 47, 2068 (1993).

\bibitem{He_PhysRevC.54.857}
Y. B. He and W. Q. Chao and C. S. Gao and  X. Q. Li,
Phys. Rev. C 54, 857 (1996).

\bibitem{Maruyama_PhysRevD.76.123015}
T. Maruyama and S. Chiba and H. J. Schulze and T. Tatsumi,
Phys. Rev. D 76, 123015 (2007).

\bibitem{GL2021prc103}
G. Lugones and A. G. Grunfeld,
Phys. Rev. C 103, 035813 (2021).

\bibitem{GL2021prd104}
G. Lugones and A. G. Grunfeld,
Phys. Rev. D 104, L101301 (2021).

\bibitem{Shao_PhysRevD.85.114017}
G. Y. Shao and M. Colonna and M. Di Toro and B. Liu and F. Matera,
Phys. Rev. D 85, 114017 (2012).

\bibitem{Mizher2010_PRD82-105016}
A. J. Mizher and M. N. Chernodub and E. S. Fraga,
Phys. Rev. D 82, 105016 (2010).

\bibitem{Li2016_PRD93-054005}
 C. F. Li and L. Yang and X. J. Wen and G. X. Peng,
Phys. Rev. D 93, 054005 (2016).

\bibitem{Maruyama2019_JCP}
T. Maruyama and T. Tatsumi,
JPS Conf. Proc. 26, 024020 (2019).

\bibitem{Gorenstein_PhysRevD.52.5206}
M. I. Gorenstein and S. N. Yang,
Phys. Rev. D 52, 5206 (1995).

\bibitem{Peshier_PhysRevC.61.045203}
A. Peshier and B. K\"ampfer and G. Soff,
Phys. Rev. C 61, 045203 (2000).

\bibitem{Li_PhysRevD.99.043001}
B. L. Li and Z. F. Cui and Z. H. Yu and Y. Yan and S. An and H. S. Zong,
Phys. Rev. D 99, 043001 (2019).

\bibitem{Zhang_PhysRevD.103.103021}
Z. Zhang and P. C. Chu and X. H. Li and H. Liu and X. M. Zhang,
Phys. Rev. D 103, 103021 (2021).

\bibitem{Xu_2003}
R. X. Xu,
Astrophysi. J. 596, L59 (2003).

\bibitem{Shi_2003}
Y. Shi and R. X. Xu,
Astrophysi. J. 596, L75 (2003).

\bibitem{Xu_2010}
R. Xu,
Int. J. Mod. Phys. D 19, 1437 (2010).

\bibitem{Freedman_PhysRevD.16.1130}
B. A. Freedman and L. D. McLerran,
Phys. Rev. D 16, 1130 (1977).

\bibitem{Baluni_PhysRevD.17.2092}
V. Baluni,
Phys. Rev. D 17, 2092 (1978).

\bibitem{Fraga_PhysRevD.63.121702}
E. S. Fraga and R. D. Pisarski and J. Schaffner-Bielich,
Phys. Rev. D 63, 121702(R) (2001).

\bibitem{Peng_EL.72.69}
G. X. Peng,
Europhys. Lett. 72, 69 (2005).

\bibitem{Xu_PhysRevD.92.025025}
J. F. Xu and G. X. Peng and F. Liu and D. F. Hou and L. W. Chen,
Phys. Rev. D 92, 025025 (2015).

\bibitem{Xu_PhysRevD.96.063016}
J. F. Xu and Y. A. Luo and L. Li and G. X. Peng,
Phys. Rev. D 96, 063016 (2017).

\bibitem{Peng_PhysRevC.62.025801}
G. X. Peng and H. C. Chiang and B. S. Zou and P. Z. Ning and S. J. Luo,
Phys. Rev. C 62, 025801 (2000).

\bibitem{Bao_EPJA.38.287}
T. Bao and G. Z. Liu and E. G. Zhao and M. F. Zhu,
Eur. Phys. J. A 38, 287 (2008).

\bibitem{Isayev_2013}
A. A. Isayev and J. Yang,
J. Phys. G: Nucl. Part. Phys. 40, 035105 (2013).

\bibitem{Chu_2018}
P. C. Chu and X. H. Li and H. Y. Ma and B. Wang and Y. M. Dong and X. M. Zhang,
Phys. Lett. B 778, 447 (2018).

\bibitem{Chen_PhysRevD.105.014011}
H. M. Chen and C. J. Xia and G. X. Peng,
Phys. Rev. D 105, 014011 (2022).

\bibitem{Chen_CPC.46.055102}
H. M. Chen and C. J. Xia and G. X. Peng,
Chin. Phys. C 46, 055102 (2022).

\bibitem{PDG2022}
R. L. Workman et al, [Particle Data Group 2022]
Prog. Theor. Exp. Phys. 2022, 083C01 (2022).

\bibitem{Lu2016}
Z. Y. Lu and G. X. Peng and S. P. Zhang and M. Ruggieri and V. Greco,
Nucl. Sci. Tech 27, 148 (2016).

\bibitem{Chen2012}
S. W. Chen and L. Gao and G. X. Peng,
Chin. Phys. C 36, 947 (2012).

\bibitem{Borsanyi.Endrodi.ea1-22JHEP}
Sz. Bors{\'a}nyi, and G. Endr{\H{o}}di and Z. Fodor and S. D. Katz and K. K. Szabo,
J. High. Energy. Phys. 01, 138 (2012).

\bibitem{Chen2022IJMPE}
H. M. Chen and L. M. Liu and J. T. Wang and M. Waqas and G. X. Peng,
Int. J. Mod. Phys. E 31, 2250016 (2022).

\bibitem{Hagedorn1980PLB}
R. Hagedorn and J. Rafelski,
Phys. Lett. B 97, 136 (1980).

\bibitem{Wen2007CITP}
X. J. Wen and G. X. Peng and P. N. Shen,
Commun. Theor. Phys. 47, 78 (2007).

\bibitem{Bhattacharyya2016EPL115}
 A. Bhattacharyya and S. K. Ghosh and R. Ray and S. Samanta,
Europhys. Lett. 115, 62003 (2016).

\bibitem{Bugaev2000PLB485}
 K. A. Bugaev and M. I. Gorenstein and H. Stoecker and W. Greiner,
Phys. Lett. B 485, 121 (2000).

\bibitem{Pal _NPA.A1010.122177}
S. Pal and A. Bhattacharyya and R. Ray,
Nucl. Phys. A1010, 122177 (2021).

\bibitem{Bhagyarathi _PhysRevD.108.074028}
B. Sahoo and K. K. Pradhan and D. Sahu and R. Sahoo,
Phys. Rev. D 108, 074028 (2023).

\bibitem{Vovchenko _PhysRevLett.118.182301}
 V. Vovchenko and M.I. Gorenstein and H. Stoecker,
Phys. Rev. Lett. 118, 182301 (2017).
	
\end{thebibliography}
\end{document}